\documentclass[submission,copyright,creativecommons]{eptcs}
\providecommand{\event}{LAFM 2013} 
\usepackage{breakurl}             
\usepackage{amssymb}
\usepackage{graphicx}
\usepackage{amsmath}
\usepackage{amsfonts}
\usepackage{stmaryrd}
\usepackage[noload]{qtree}
\usepackage[all]{xypic}

\def\per#1{\textsf{P}(#1)}
\def\wper#1{\textsf{P}_{\mathsf{w}}(#1)}

\def\univ{\textbf{U}}
\def\equivdef{\buildrel \mathsf{def} \over \Longleftrightarrow}
\def\eqdef{\buildrel \textsf{def} \over =}
\def\relatedto#1{\buildrel #1 \over \rightarrow}

\def\event{\textsf{F}}

\newtheorem{definition}{Definition}
\newtheorem{theorem}{Theorem}
\newtheorem{corollary}{Corollary}

\title{Automated Reasoning over Deontic Action Logics with Finite Vocabularies}
\author{Pablo F. Castro
\institute{CONICET\\ Argentina}
\institute{Departamento de Computaci\'on \\
Universidad Nacional de R\'io Cuarto\\
 R\'io Cuarto,  Argentina.
}
\email{pcastro@dc.exa.unrc.edu.ar}
\and
Thomas S.~E.~Maibaum
\institute{Department of Computing \& Software\\
McMaster University\\
Hamilton (ON), Canada.}
\email{tom@maibaum.org}
}
\def\titlerunning{Automated Reasoning over Deontic Action Logics with Finite Vocabularies}
\def\authorrunning{Pablo F.~Castro \& Thomas S.~E.~Maibaum}
\begin{document}
\maketitle

\begin{abstract}
In this paper we investigate further the tableaux  system for a deontic action logic we presented in previous work. This tableaux system uses atoms (of a given boolean algebra of action terms) as labels of formulae. This allows us to embrace parallel execution of actions and action complement, two action operators that may present difficulties in their treatment. One of the restrictions of this logic is that it uses vocabularies with a finite number of actions. In this article we prove that this restriction does not affect the coherence of the deduction system; in other words, we prove that the system is complete with respect to language extension. We also study the computational complexity of this extended deductive framework and we prove that the complexity of this system is in PSPACE, which is an improvement with respect to related systems.
\end{abstract}

\section{Introduction}

Tableau systems \cite{Smullyan68, Gore95, Fitting72} are practical proof systems that are representative of an important stream of research in  automated theorem proving \cite{Fitting}. The basic idea behind these kinds of proof systems is proving by refutation, i.e., to prove a formula $\varphi$, we start with $\neg \varphi$ and then we try to derive a contradiction using the rules provided by the logical system.  Usually, if a formula is not provable,  we get a counterexample (a model which satisfies the negation of the formula). Several tableau systems have been proposed for logics used in computer science, some examples are: \textit{dynamic logics} \cite{Pratt78,Giacomo-Massacci}, \textit{modal  logics} \cite{Fitting72,Massacci2000} and \textit{temporal logics} \cite{Gore95, EmersonSurvey}.
	
	In \cite{DEON08} we introduced a tableaux method for the deontic action logic presented in \cite{JAL2009}; this logic is a \textit{modal action logic} \cite{MaibaumKhosla} which uses boolean operators on actions (parallel execution and complement of actions) as well as the standard deontic predicates over actions, i.e., \textit{permission}, \textit{obligation} and \textit{prohibition}. We have proposed this logic for reasoning about  fault-tolerant programs \cite{JAL2009}; the deontic predicates seem suitable to formalize concepts such as \textit{fault}, \textit{violation}, and \textit{fault-recovery}. Some intuitions and examples of the use of this logic in specification and verification of fault-tolerant systems are shown in \cite{PhDThesis}. We believe that tableau methods can help us provide automated theorem provers for this logic, enabling automation of the analysis of software specifications. 	
	
	In this paper we extend the system presented in \cite{DEON08} and present further results. In particular, we redefine the rules of the system in such a way that the method for checking formula validity is in PSPACE, and we prove that the restriction to finite action vocabularies does not affect the completeness of the method. This was not proven in the paper cited above.	In technical words,  we consider a finite number of actions in vocabularies; this implies that extending the vocabularies may affect the validity of formulae, for instance, we may add some extra actions in a vocabulary that may introduce new scenarios in a model that could falsify a formula that has been proven valid for the original vocabulary. We provide a formal machinery to tackle this problem; corollary \ref{theorem:bound} of section \ref{section:language-extension} states that there is a bound on the number of actions to be considered when proving the validity of formulae. More precisely, given a formula, we can calculate the number of actions that we must consider in the vocabulary to prove its ``global'' validity (i.e., its validity in every vocabulary).	Having a finite number of actions has some theoretical benefits, for instance, this implies that the underlying algebra of action is atomic, and these atoms can be used as labels of formulae to build canonical models. This also implies that the logic is compact in contrast with similar logics. We discuss this in section \ref{section:related}.	Furthermore, in section \ref{section:complexity} we study the complexity of this tableaux system and we show that the system is in PSPACE, that is, it is aligned with the complexity of most modal logics. Interestingly, most of the  dynamic logics with boolean operators proposed in the literature are EXPTIME-Complete. 
  
	The paper is organized as follows. In the next section we give a brief description of the deontic action logic. In section \ref{section:tableaux} we introduce the tableaux system for  this logic together with the basic properties of this system.  In section \ref{section:language-extension} we prove some theorems about how formula validity is preserved when  the vocabulary is extended. Finally, in section \ref{section:complexity} we investigate the complexity of the methods proposed below, and then we discuss some conclusions. 

\section{Background}\label{section:background}
	Deontic action logics \cite{J.J.Meyer,BroersenThesis,MaibaumKhosla} (also called dynamic deontic logics)  are modal action logics \cite{MaibaumKhosla, DynamicLogic} with deontic predicates over actions. These logics can be classified as ``\textit{ought-to-do}'' deontic logics, since the deontic operators are applied to actions, in contrast to ``\textit{ought-to-be}'' logics where deontic operators  are applied to predicates. For example, the standard deontic system KD \cite{Chellas} is an ``ought-to-be'' deontic logic. In this section we describe, briefly, the deontic logic  (called DPL) presented in \cite{JAL2009};  for further details, the reader is referred to that paper.

	The language of the logic is given by a vocabulary $V= \langle \Delta_{0}, \Phi_{0} \rangle$, where $\Delta_{0}$ is a
finite set of primitive actions (denoted by  $a, b, c, d, ...$), and a set $\Phi_{0}$ of
primitive propositions (denoted by $p, q, s, \dots$). Using these two sets  one can build complex formulae by employing  the modal connectives, the deontic predicates and the boolean operators over actions. 
	The set $\Delta$ of action terms and  the set  $\Phi$ of formulae over $V$ are described by the following grammars: 
\begin{center}
\begin{tabular}{p{10cm}}
\\
$\alpha ::= a_i \mid \alpha \sqcap \alpha \mid \alpha \sqcup \alpha \mid \overline{\alpha} \mid \emptyset \mid \univ$ \\
$\varphi ::= p_i \mid \neg \varphi \mid \varphi \rightarrow \varphi \mid \per{\alpha} \mid \wper{\alpha} \mid [\alpha]\varphi \mid \alpha_1 =_{act} \alpha_2$\\
\\
\end{tabular}
\end{center}
where $a_i \in \Delta_0$. $\alpha_1 \sqcup \alpha_2$ is the non-deterministic choice between actions $\alpha_1$ and 
$\alpha_2$, $\alpha_1 \sqcap \alpha_2$ is the parallel execution of actions $\alpha_1$ and $\alpha_2$, $\overline{\alpha}$ 
denotes the execution of an alternative action to $\alpha$, $\emptyset$ is an impossible action and $\univ$ denotes 
the execution of any action. 	
	In addition to the standard boolean connectives, we have the following formulae: $[\alpha]\varphi$ asserts that after any execution  
of $\alpha$, $\varphi$ is true; and $\alpha_1 =_{act} \alpha_2$ states that actions $\alpha_1$ and $\alpha_2$ are equal.
 	We consider two permission predicates: $\wper{\alpha}$ is called \textit{weak} permission; it is true  
when $\alpha$ is allowed to be executed in some scenarios. On the other hand, $\per{\alpha}$ is called \textit{strong} permission; this 
formula is true when $\alpha$ is permitted to be executed in any scenario. The two versions of permission can be found in the deontic literature. Using permissions, we introduce other deontic operators such as obligation, prohibition, etc.
	Note that in this logic the interpretation of deontic predicates is 
independent of the modal operators (they have a different interpretation in the semantic structures), whereas in related work (e.g., \cite{J.J.Meyer,BroersenThesis})  
the deontic operators are reduced to modal formulae.

	Let us briefly introduce the semantics of the logic with some remarks:
\begin{definition}[structures] \label{def:models} Given a vocabulary $V = \langle \Phi_{0}, \Delta_{0} \rangle$, a \textit{V-Structure} is a tuple 
$M  = \langle \mathcal{W}, \mathcal{R}, \mathcal{E}, \mathcal{I}, \mathcal{P} \rangle$ where: 
\begin{itemize}
        \item $\mathcal{W}$ is  a set of worlds; 
        \item $\mathcal{R}$ is an $\mathcal{E}$-labeled relation between worlds, 
        s.t., if $(w,w',e) \in \mathcal{R}$ and $(w,w'',e) \in \mathcal{R}$, then $w'=w''$;
       \item $\mathcal{E}$ is a non-empty set of (names of) events. 
       \item  $\mathcal{P} \subseteq \mathcal{W} \times \mathcal{E}$ is a relation which indicates which event is permitted in which world; 
       \item $\mathcal{I}$ is a function s.t. for every $p \in \Phi_{0}: \mathcal{I}(p) \subseteq \mathcal{W}$ and for every $\alpha \in \Delta_{0}: \mathcal{I}(\alpha) \subseteq \mathcal{E}$.             
\end{itemize}
$\mathcal{I}$ has to satisfy the following properties:
              \begin{description}        
                        \item[I.1] \label{cond:int1} For every $\alpha_{i} \in \Delta_{0}$: 
                        $|\mathcal{I}(\alpha_{i}) - \bigcup\{\mathcal{I}(\alpha_{j}) \mid \alpha_{j} \in (\Delta_{0} - \{\alpha_{i}\})\}| \leq 1$.
                        \item[I.2] \label{cond:int2} For every $e \in \mathcal{E}$: if $e \in \mathcal{I}(\alpha_{i})\cap \mathcal{I}(\alpha_{j})$, 
                         where $\alpha_{i} \neq \alpha_{j} \in \Delta_{0}$, then: \\
                        $\cap\{ \mathcal{I}(\alpha_{k}) \mid \alpha_{k} \in \Delta_{0} \wedge e \in \mathcal{I}(\alpha_{k})\} = \{e\}$.
                        \item[I.3] \label{cond.int3} $\mathcal{E} = \bigcup_{\alpha_{i}\in \Delta_{0}} \mathcal{I}(\alpha_{i})$.
              \end{description}  
\end{definition}	
\noindent	We can extend the function $\mathcal{I}$ to well-formed formulae and action terms as follows:
\begin{itemize}
	\item  $\mathcal{I}(\neg \varphi) \eqdef \mathcal{W} - \mathcal{I}(\varphi)$, 
	\item  $\mathcal{I}(\varphi \rightarrow \psi) \eqdef \mathcal{I}(\neg \varphi) \cup \mathcal{I}(\psi)$, 
	\item $\mathcal{I}(\alpha \sqcup \beta) \eqdef \mathcal{I}(\alpha) \cup \mathcal{I}(\beta)$, 
	\item $\mathcal{I}(\univ) \eqdef \mathcal{E}$,
	\item  $\mathcal{I}(\alpha \sqcap \beta) \eqdef \mathcal{I}(\alpha) \cap \mathcal{I}(\beta)$, 
	\item $\mathcal{I}(\overline{\alpha}) \eqdef \mathcal{E} - \mathcal{}I(\alpha)$, 
	\item  $\mathcal{I}(\emptyset) \eqdef \emptyset$.
\end{itemize}
	Note that here we do not follow the traditional approach of interpreting each action as
a relation (e.g., see \cite{BooleanModalLogic}); instead we interpret each action as a set of ``\textit{events}'', the events in which it ``\textit{participates in}'' during its execution.  
 Then, the action combinators are interpreted as the classical boolean set operators. Note that the restrictions on models 
(\textbf{I.1}  and \textbf{I.2}) imply that we have one point sets in the family of the event sets, i.e., intuitively every ``event'' is 
produced by a combination of actions in our systems (system actions and environmental actions). (So, events can be seen as collections of actions
 to be executed in parallel.)  Then, if we take a maximal set of
actions, the execution of this set only produces an event in our system; in other words, this set of actions is complete in the sense 
that they describe unambiguously one event in the system execution.  We have presented a sound and complete axiomatic system 
for this logic in \cite{JAL2009}.
Given a structure $M  = \langle \mathcal{W}, \mathcal{R}, \mathcal{E}, \mathcal{I}, \mathcal{P} \rangle$, we define the relation $\vDash$ between worlds, models, and formulae, as follows:
\begin{itemize}
	\item $w, M \vDash p \equivdef w \in \mathcal{I}(p)$.
	\item $w, M \vDash \varphi \rightarrow \psi \equivdef$ not $w, M \vDash \varphi$ or $w, M \vDash \psi$.
	\item $w, M \vDash \neg \varphi \equivdef$ not $w, M \vDash \varphi$.
        \item $w, M \vDash \alpha =_{act} \beta \equivdef \mathcal{I}(\alpha) = \mathcal{I}(\beta)$.
        \item $w, M \vDash [ \alpha ] \varphi \equivdef$ for all  $w' \in \mathcal{W}$ and $e \in \mathcal{I}(\alpha)$ if 
        $w \relatedto{e} w'$ then $ w', M \vDash \varphi$.
        \item $w, M \vDash \per{\alpha} \equivdef$ for all $e \in \mathcal{I}(\alpha)$, $\mathcal{P}(w,e)$ holds. 
        \item $w, M \vDash \wper{\alpha} \equivdef$ there exists some $e \in \mathcal{I}(\alpha)$ such that $\mathcal{P}(w,e)$
\end{itemize}
 	For the other standard formulae the definition is as usual. 

\subsection{Related Logics}\label{section:related}
	We make a brief digression to compare the logic presented above with related formalisms. Boolean Modal Logic \cite{BooleanModalLogic,Blackburn}  combines modal operators with boolean operators over actions. In this logic a relational semantics is provided: each action is interpreted as a relationship in the semantic structures. In this setting, the universal action is interpreted as the universal relationship between states or worlds. In the logic presented  above, the universal action is relative to the actual state, that is, this action characterizes all the reachable states from a given state, and therefore the complement is also relative to the actual state. As it is argued in \cite{BroersenThesis}, relative complements are more useful when reasoning about computer systems. To the authors' knowledge, no tableau systems have been proposed for boolean modal logic with relative complement.

 It is worth remarking that in our logic the existence of boolean atoms in the boolean algebra of actions allows us to express exactly which  actions are involved in a given transition: each atomic action denotes exactly one event. This allows us to prove the  strong completeness of the logic and therefore we get also the compactness of the axiomatic system presented in \cite{JAL2009}. BML  is not compact; this can be easily proven using the fact that the vocabularies in BML contain an infinite number of actions \cite{BroersenThesis}. 

	On the other hand, Segerberg \cite{Segerberg82} presents a deontic action logic with boolean operators where permissions are characterized as ideals of the boolean algebra of actions. As shown in \cite{TrypuzKulicki2010}, the absence of atoms in Segerberg's logic implies that the so-called closure principle (\textit{any event is allowed or forbidden}) cannot be captured by Segerberg's logic. 	In section \ref{section:complexity} we compare the time complexity of the tableaux method proposed below with the complexity of  the logics referenced in this section.

\section{A Tableaux System for DPL}\label{section:tableaux}
       In this section we describe the tableaux system introduced in \cite{DEON08}. We introduce some minor changes to the original system to be able to improve its complexity (see below); 
note that  formulae are enriched with labels; intuitively, each label indicates a state in the semantics where 
the formula is true. Labeled systems are usual for many logics and tableaux systems. An introduction to these systems can be found in \cite{Fitting72,Gabbay96}. We adapt 
these techniques to our modal action logic, showing that deontic operators fit neatly into the system; the duality between the strong and weak permissions resembles the duality between modal necessity and 
modal possibility. As shown in \cite{DEON08} this system is sound and complete.       
       
     	A labeled, or prefixed, formula has the following structure: $\sigma : \varphi$, where $\sigma$ is a label made up of a sequence
of boolean (action) terms built from a given vocabulary. We use the following notation for sequences: $\langle \rangle$ (\textit{the empty sequence}), $x \centerdot xs$ 
(\textit{the sequence made of an element $x$ followed by a sequence $xs$}); we also use the same notation 
to denote the concatenation of two sequences.

From here on we consider a fixed vocabulary: $V = \langle \Phi_{0}, \Delta_{0}\rangle$. We
denote by $\Phi_{BA}$ some complete and decidable axiomatization of boolean algebras \cite{Monk}; If 
$ \Delta_0 = \{a_1,...,a_n\}$, we add the equation $a_1 \sqcup \dots \sqcup a_n = \univ$ to the set $\Phi_{BA}$. We denote by 
$\Delta_{0}/ \Gamma$ the boolean terms over $\Delta_{0}$ modulo a set of axioms $\Gamma$; usually, 
$\Gamma$ is an extension of the theory of boolean algebras, i.e., $\Phi_{BA} \subseteq \Gamma$. We write $\Gamma \vdash_{BA} t_{1} =_{act} t_{2}$,
if the equation $t_{1} =_{act} t_{2}$ is provable from the boolean theory $\Gamma$ using equational calculus. This implies that our 
method depends on some suitable method to decide boolean algebras. Using this notation, we denote by 
$At(\Delta_{0} / \Gamma )$ the set of atoms in the boolean algebra of terms modulo $\Gamma \cup \Phi_{BA}$ (note that the boolean
algebra is atomic because the set of primitive action symbols is finite). In the same way, we denote by $At_{\sqsubseteq \alpha}(\Delta_{0}/\Gamma)$ 
the set of atoms $\gamma \in At(\Delta_{0}/\Gamma)$ such that $\Gamma \vdash_{BA} \gamma \sqsubseteq \alpha$, where $\sqsubseteq$ is the order relation of the algebra,  
and $At_{\sqsubset \alpha}(\Delta_{0} / \Gamma)$ denotes the strict version of this set.

A \emph{tableau} is an (n-ary) rooted tree where nodes are labeled with prefixed formulae,
and a \emph{branch} is a path from the root to some leaf.
    Intuitively, a branch is a tentative model for the initial formula (which we are trying to prove valid). Given a 
 branch $\mathcal{B}$, we denote by $EQ(\mathcal{B})$ the equations appearing in $\mathcal{B}$.

\begin{figure}[h]
\centering
\includegraphics[scale=1]{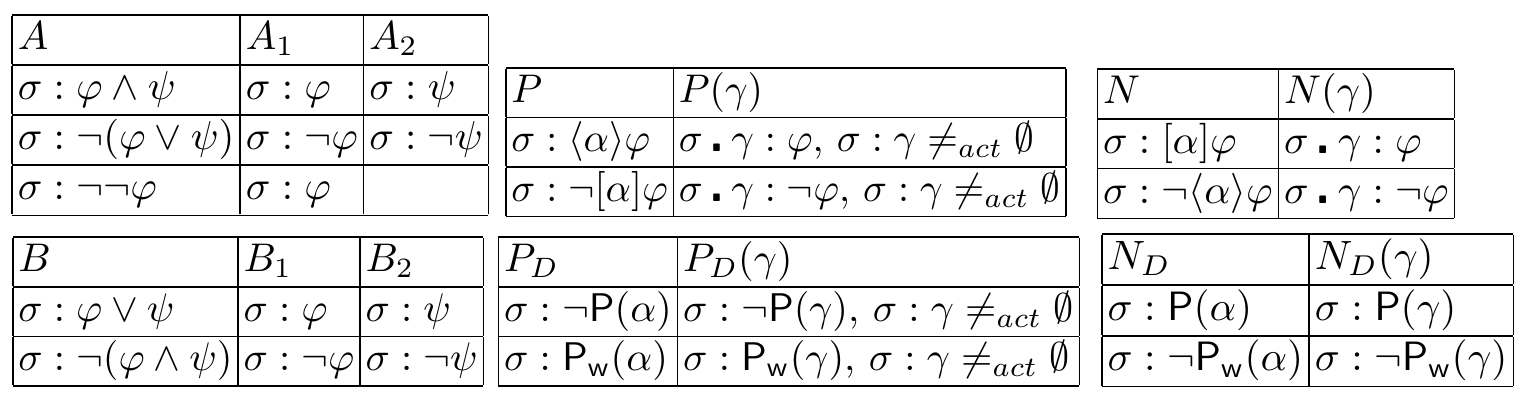}
\caption{Classification for deontic formulae.}\label{fig:deontic-notation} 		
\end{figure}

\begin{figure}[t]
\centering
\includegraphics[scale=1]{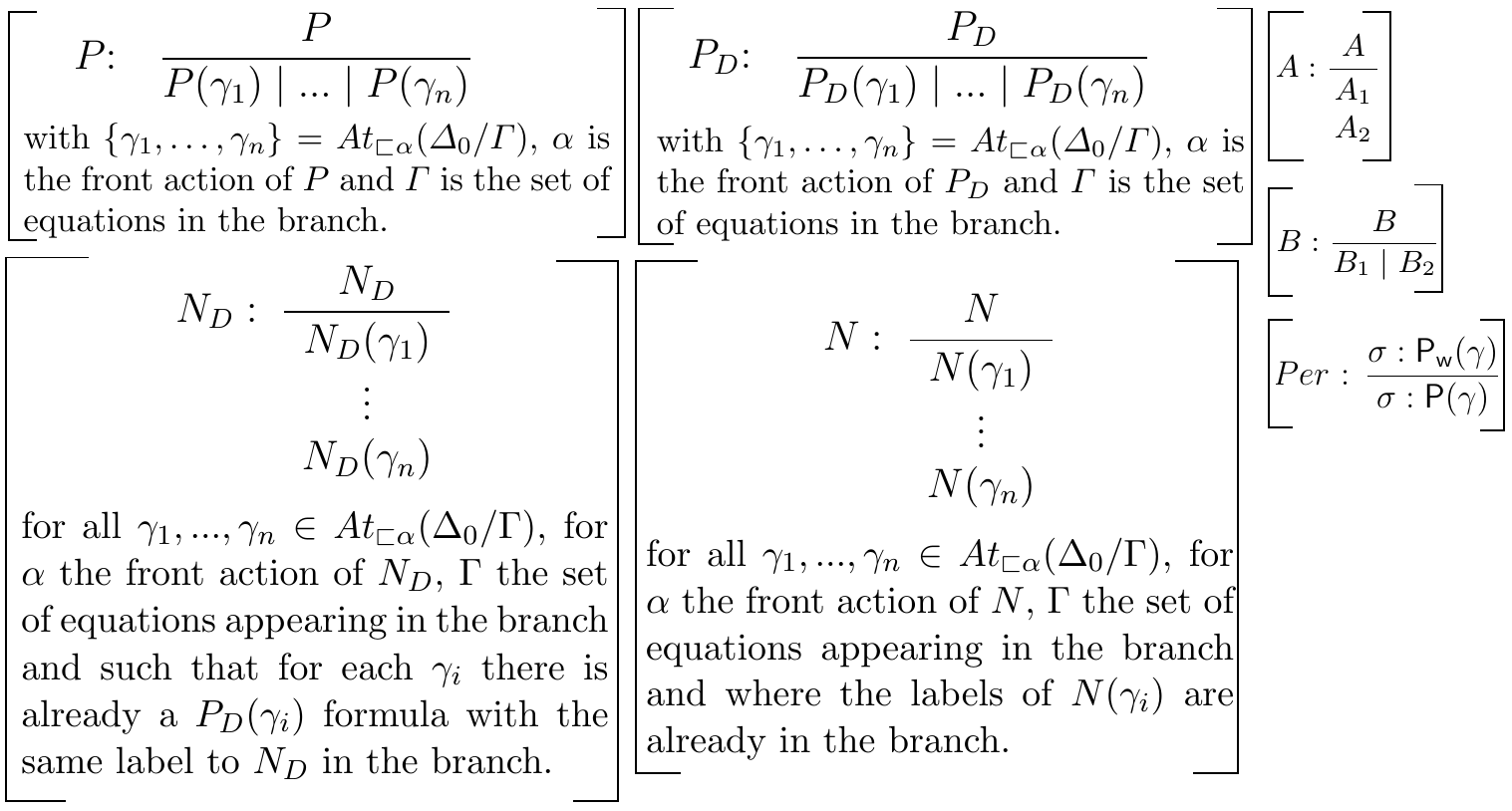}
\caption{Tableau Rules}\label{fig:rules}
\end{figure}

  In figure \ref{fig:deontic-notation} we introduce a classification of formulae which 
is useful for presenting the rules of the tableaux calculus (see figure \ref{fig:rules}).   
Standard propositional formulae are classified following Smullyan's unifying notation \cite{Smullyan68}.    
	We also introduce the less standard classification for  modal logics. (We follow the standard notation for modal logics \cite{Fitting72}.)  
For each prefixed formula of type $P$ or $N$, we define formulae $P(\gamma)$ and $N(\gamma)$, respectively. 
Here $\gamma$ is some action term which is needed to define 
these formulae (see the rules below). Note that for any formula $P$, $P(\gamma)$ denotes two formulae. Finally, we introduce a new classification for deontic formulae (formulae $P_D$ and $N_D$). 
Although the deontic operators are, in some sense, similar to the modal operators, we need to distinguish them; the deontic predicates state properties about transitions, whereas  
the modal operators state properties about states related to the actual state.
	Using the above classification of formulae, we can introduce the rules of the tableaux method. In the definition 
of these rules we use \textit{front action} of a $P$, $N$, $P_D$ or $N_D$ formula to refer to the action nearest the root in the syntax 
tree corresponding to this formula.

	 The rules of this calculus can be found in figure \ref{fig:rules}. The rules for standard boolean operators are as usual. Rule $N$ is standard for \textbf{K} modal logics \cite{Fitting72}; 
it does not introduce new labels in the branch, but it adds new formulae to labels already in the branch; intuitively, for all (the states denoted by) the labels reachable from the current state, the $N$ formula must be true. 
The rule for deontic necessity is similar, but it adds the corresponding deontic formulae to labels already in the branch and for which there is a $P_D$ formula with the same action in the branch.

	Notice the rules  $P$ and $P_D$ for modal and deontic possibility, respectively;  given a $P$ formula, rule $P$ creates one branch for each possible execution of the front action in the formula; although the rule for deontic 
possibility is very similar, note that deontic possibility does not create new labels, because permissions only predicate over transitions. Note that, in these rules, an inequation saying that the action must not be impossible is added 
in each branch, allowing us to avoid adding labels that cannot exist in the semantics. Finally, rule $Per$ states that, if an action which is atomic (in the sense that it cannot have different executions, i.e., not participate in different events) is
weakly allowed, then it is also strongly allowed. 
	
 	We do not state any rule for equality;  
this is because equality reasoning is implicit in our calculus (see below the definition of boolean closed). 
For simplicity of the presentation of the concepts, we rule out those formulae of the form: $[\alpha](\alpha =_{act} \beta)$. This does not affect the completeness of the method since formulae of this
kind are equivalent to formulae where equations do not appear after modalities \cite{JAL2009}.
%
  Let us introduce the notions of \textit{closed}, \textit{boolean closed}, \textit{deontic closed} and \textit{open branch}. Keep in mind that a branch is a set of prefixed formulae.	

Given a branch $\mathcal{B}$ and a boolean theory $\Gamma$, we say that
$\mathcal{B}$ is \emph{deontic closed} with respect to $\Gamma$ if it satisfies at least one of the
following conditions: (i) $\sigma:\per{\alpha} \in \mathcal{B}$ and $\sigma:\neg \per{\alpha} \in \mathcal{B}$, for some label $\sigma$;
	(ii) $\sigma:\wper{\alpha} \in \mathcal{B}$ and $\sigma:\neg \wper{\alpha} \in \mathcal{B}$, for some label $\sigma$;	
	(iii) $\sigma:\neg \per{\alpha} \in \mathcal{B}$ and $\sigma:\wper{\alpha} \in \mathcal{B}$, for some label $\sigma$.

	Note that we have not included $\sigma:\per{\alpha}$ and $\sigma:\neg \wper{\alpha}$ as being mutually contradictory; 
this is because they are not contradictory when $\Gamma \vdash_{BA} \alpha =_{act} \emptyset$. This fact yields 
the definition of \emph{extended boolean theory}: 
\[
\textstyle
	EQ^{*}(\mathcal{B}) = \{ (\alpha \sqcap \beta =_{act} \emptyset) \mid \sigma:\per{\alpha}, \sigma:\neg \wper{\beta} \in \mathcal{B} \} \cup EQ(\mathcal{B})
\]
It is useful for us to introduce
the notion of \textit{boolean closed} branch; intuitively these branches are inconsistent boolean theories. 
A branch $\mathcal{B}$ is \emph{boolean closed} iff $EQ^{*}(\mathcal{B}) \vdash_{BA} \emptyset =_{act} \univ$,
or $EQ^{*}(\mathcal{B}) \vdash_{BA} \alpha = _{act} \beta$ and $\alpha \neq_{act} \beta \in \mathcal{B}$.

	Finally, we say that a branch is \emph{closed} if either it contains a labeled propositional variable $\sigma:p$
and a labeled negation of it $\sigma:\neg p$, or it is deontic closed or boolean closed. Note that rule $N_D$ only takes into account labels 
that contain a $P_D$ formula. In \cite{DEON08} this rule creates a labelled deontic formula for each atom of the corresponding action, implying that in that system 
the space needed in a branch is exponential w.r.t. formula length.

\section{Completeness with respect to language extension}\label{section:language-extension}
	Recall that  vocabularies contain a finite number of primitive actions. This is different from what happens  in 
dynamic logics, where vocabularies have an infinite number of actions. This assumption allows us to obtain atoms in the corresponding 
boolean algebra of action terms. These atoms are useful for proving completeness and compactness. However, doing this we also 
constrain our deductive machinery to only take into account a restricted number of actions. Sometimes, we will be interested in proving 
properties that are valid in every vocabulary, and not only in a particular one. This extension idea then represents situations where we have embedded our component in a
 larger one with a set of primitive actions extending those of the component. 

 Consider the formula: $\langle \alpha \rangle \varphi \rightarrow [ \alpha ] \varphi$. 
	 This formula is not valid, but if we build the tableau for it considering a vocabulary with $a$ as the unique action, the final tree has no open branches. 
This only shows that this formula is valid for a vocabulary with  one primitive action. 
 Intuitively, if we think of a theory as a specification of a computing system, 
we take the view that the vocabulary describes all the actions that can be executed during a run of the system being specified.  Adding more actions to the 
vocabulary can be understood as incorporating new behavior to the system, or taking into account more actions from the environment (e.g., some additional 
interaction with the users).  Sometimes, we will be interested in proving that some properties are valid in any vocabulary. This has the obvious interpretation that 
these are properties that are valid for a system and any extension of it.

	Summarizing, our specification only gives us a partial picture of a system. Because of this, system properties are hard to 
verify (using tableaux or other formal systems). After all, perhaps we may not be taking into account some actions important for the 
property to be proven.
        The following theorems give us some machinery to address this difficulty. Corollary \ref{theorem:bound} says that we can verify
a property (using tableaux) restricting our attention  to a finite number of actions; if for this number of actions, this formula
is valid, then it will be valid for any language extension (containing, potentially, any number of actions). 
Some auxiliary notions are needed and we introduce the concepts of \textit{normal form}, \textit{disjunctive normal form},
and \textit{existential degree}, and then we present the theorems.

        The \textit{degree} of a formula $\varphi$ (denoted by $\mathsf{d}(\varphi)$) is the length of the longest string of nested modalities (taking permission
as being of degree $0$). For any formula $\varphi$ we denote by $Pr(\varphi)$ the set of primitive actions appearing in $\varphi$. Given a vocabulary $\langle \Delta_{0}, \Phi_{0}\rangle$, we adapt the definition of \textit{normal form of degree n} 
given in \cite{Fine75} to our logic. We denote by $F_i$ the set of formulae of normal form of degree $i$, defined as follows:
\begin{itemize}
        \item $F_{0}$ is the set of formulae of the form $*\varphi_{1} \wedge... \wedge *\varphi_{h}$, where for each $i$: $\varphi_{i} \in \Phi_{0}$ or 
        $\varphi_i$ is a deontic predicate, and $*$ is $\neg$ or blank.
        \item $F_{n+1}$ is the set of formulae of the form: $\theta \wedge  *\langle \alpha_{1} \rangle \varphi_{1} \wedge ....\wedge * \langle \alpha_{k} \rangle \varphi_{k}$,  
        where $\theta \in F_{0}$, $\varphi_{i} \in F_{n}$ for all $1 \leq i \leq k$, $*$ is $\neg$ or blank. 
        ($\theta$ may not appear in the formula, in which case we only consider everything but not $\theta$.)   
\end{itemize}
The set of normal form formulae is $F = \bigcup_{i=1}^{\infty} F_{i}$. If a formula is in normal form, we say that it is a 
NF formula. Any formula of degree $\leq n$ is equivalent to $\bot$ or a disjunction of normal forms 
of degree $n$:
\begin{theorem} \label{theorem:NF} Given a vocabulary $V$ and formula $\varphi$ of degree $\leq n$, either there exist  NF formulae $\phi_i$ of degree $n$, such that 
$w, M \vDash \varphi \leftrightarrow \bigvee \varphi_i$ for any $V'$-structure  $M$ (with $V \subseteq V'$); or there is no $V'$-structure $M$ s.t. $w, M\vDash \varphi$.
\\
\textbf{Proof} See the proof given in \cite{Fine75} and use the property
 $\vdash \langle \alpha \rangle (\varphi \vee \psi) \leftrightarrow \langle \alpha \rangle \varphi \vee \langle \alpha \rangle \psi$, which is valid in any vocabulary.
\end{theorem}
	Note that, in general, a NF formula can be expressed using the following schema:
\[\textstyle
	\theta \wedge \Delta \wedge \bigwedge^n_{i=1}\langle \alpha_i \rangle \varphi \wedge \bigwedge^m_{j=1} \neg \langle \beta_j \rangle \psi
\] 
where $\theta$ is a conjunction of propositional variables or negations of them, and $\Delta$ is a conjunction of deontic predicates or negations  
of them.

        If a formula is a disjunction of normal forms, we say that this formula is in \textit{disjunctive normal form} (or DNF for short).
We call $\wper{\alpha}$, $\neg \per{\alpha}$ and $\langle \alpha \rangle \varphi$ \textit{existential formulae}, i.e., \textit{existential 
formulae} are those whose semantics is given in terms of an existential quantifier. 
	Given a NF formula $\varphi$, with $\textsf{d}(\varphi)=n$, we can define a set of formulae $\textsf{SF}(\varphi, k)$,
for every $k \leq n$, called the \textit{subformulae at level} $k$. For $k=0$ we define:
\begin{itemize}
	\item If $ \varphi = \bigwedge^n_{i=1} * p_i$, i.e., it is a conjunction of propositions or 
	negations of them, then for this case the definition is:
	$
		\textsf{SF}(\bigwedge^n_{i=1} * p_i, 0) = \bigcup^n_{i=1}\{ *p_i \} 
	$
	\item If $\varphi = \bigwedge^m_{j=1}*\per{\alpha_j} \wedge \bigwedge^t_{k=1}*\wper{\beta_k}$, i.e., the formula is a conjunction 
	of deontic formulae or negations of them, then we define:
	\[\textstyle
		\textsf{SF}(\bigwedge^m_{j=1}*\per{\alpha_j} \wedge \bigwedge^t_{k=1}*\wper{\beta_k}, 0) = \bigcup^m_{j=1}\{* \per{\alpha_j}\} \cup \bigcup^t_{k=1} \{* \wper{\beta_k}\}
	\]
	\item In the case of a conjunction of propositional formulae and deontic formulae we can use the two definitions above, that is:
	$
		\textsf{SF}(\theta \wedge \Delta, 0) = \textsf{SF}(\theta, 0) \cup \textsf{SF}(\Delta, 0)
	$
	\item In the general case, we define:\\
\hspace*{-5mm} {\small{$\displaystyle
		\textsf{SF}(\theta \wedge \Delta \wedge (\bigwedge^n_{i=1}\langle \alpha_i \rangle \varphi_i) \wedge (\bigwedge^m_{j=q} \neg \langle \beta_j \rangle \psi_j), 0) =
		\textsf{SF}(\theta) \cup \textsf{SF}(\Delta) \cup \bigcup^n_{i=1} \{ \langle \alpha_i \rangle \varphi_i \} \cup \bigcup^m_{j=1} \{ \neg \langle \beta_j \rangle \psi_j \}
	$}}	
\end{itemize}
	For the case of $k>0$, we define:
$$
\textstyle
\textsf{SF}(\theta \wedge \Delta \wedge (\bigwedge^n_{i=1}\langle \alpha_i \rangle \varphi_i) \wedge (\bigwedge^m_{j=q} \neg \langle \beta_j \rangle \psi_j), k+1) =
		\bigcup^n_{i=1} \textsf{SF}(\varphi_i, k) \cup \bigcup^m_{j=1} \neg \textsf{SF}(\psi_j, k)
$$

\noindent where given a set $S$ of formulae, we denote by $\neg S$, the set containing the negations of the formulae in $S$. (We also suppose 
that several negations over a formula are simplified, i.e., instead of having $\neg \neg p$ we have $p$.) In some sense, the set 
$\textsf{SF}$ indicates which set of subformulae must be true at a given level. We use $\#_{\exists} S$ to denote the number of 
existential formulae in the set $S$. Using this definition, we can define the \textit{existential degree} of a NF formula 
$\varphi$, denoted by $\textsf{D}_{\exists}$, which is defined as follows:
$
	\textsf{D}_{\exists}(\varphi) = \textsf{max}_{ 0 \leq i \leq n}\{\#_{\exists} \textsf{SF}(\varphi, i)\}.
$
	We can extend this definition to DNF formulae, as follows:
$
	\mathsf{D}_{\exists}(\varphi_{1} \vee ... \vee \varphi_{k}) = max\{\mathsf{D}_{\exists}(\varphi_{1}),..,\mathsf{D}_{\exists}(\varphi_{k})\}
$

	Note that  function $\mathsf{D}_\exists$ can be extended to cope with any formula: to obtain $\mathsf{D}_\exists(\varphi)$ (where $\varphi$ may not be a 
DNF formula) calculate the maximum number of existential subformulae in the same level of the syntax tree of such a formula. Note that this number 
coincides with the existential degree of an equivalent DNF formulae (obtained by using theorem \ref{theorem:NF}).

	The idea is to use the sets $\textsf{SF}$ to define smaller models of $\varphi$. First, we need to define the notion 
of $n$-reachable. Given a model $M$ and a state $w$, we say that a state $v$ is $n$-reachable (or reachable in $n$-steps) from 
$w$, if there exists a path $w \relatedto{e_1} w_2 \relatedto{e_2} \dots \relatedto{e_n} v$ in $M$. Note that our logic 
has the \textit{unraveling property} \cite{Blackburn}, i.e., if a formula $\varphi$ is satisfiable in a model $M$ and state 
$w$, we can build a model $M'$ unraveling $M$ such that $w$ and $M'$ satisfies $\varphi$ and this new model is a tree, i.e., 
it does not have cycles. For the following results we restrict our attention to tree models; the unraveling property guarantees that these theorems  
extend to any other model.
   
	If $\textsf{d}(\varphi)=n$, then 
we can define a mapping $L_w$ from the states reachable in $M$ from $w$ in $n$ or less steps, to the subformulae of $\varphi$, as follows:
\[
	L_w(v) = \{ \psi \in \textsf{SF}(\varphi, k) \mid v \mbox{ is $k$-reachable from } w \mbox{ and } v, M \vDash \psi \mbox{ with } k \leq n \}.
\]  
	Using the definition of $L_w(v)$ we prove that, given a formula $\varphi$ and a model of this formula, we can define a new model that has 
an out-degree (the number of transitions coming out of any state) less than or equal to $\mathsf{D}_{\exists}(\varphi)$.
\begin{theorem} \label{theorem:NF-Models}	Given a NF formula $\varphi$ and a model $M=\langle \mathcal{W}, \mathcal{R}, \mathcal{E}, \mathcal{I}, \mathcal{P} \rangle$ 
over a vocabulary $V = \langle \Delta_0, \Phi_0 \rangle$, 
if $w, M \vDash \varphi$, then there exists a model $M^{\varphi}_w$ such that
$v, M^{\varphi}_{w} \vDash L_w(v)$, for every $v \in \mathcal{W}^{\varphi}_{w}$ and $M^{\varphi}_w$ has an out-degree less than or equal to $\mathsf{D}_{\exists}(\varphi)$.\\
\textbf{Proof:}
	First let us define the new model, consider a model $M= \langle \mathcal{W}, \mathcal{R}, \mathcal{E}, \mathcal{I}, \mathcal{P} \rangle$ over a vocabulary $V = \langle \Delta_0, \Phi_0 \rangle$ and a NF formula $\varphi$ (of degree $n$) such that 
$w, M \vDash \varphi$. The labeling $L$ helps us to define a new model $M^{\varphi}_{w} = \langle \mathcal{W}^{\varphi}_{w}, \mathcal{R}^{\varphi}_w, \mathcal{E}^{\varphi}_w, \mathcal{I}^{\varphi}_w, \mathcal{P}^{\varphi}_w \rangle$
as follows:
\begin{itemize}
	\item $\mathcal{E}^{\varphi}_w = \mathcal{E}$.
	\item We define $\mathcal{R}^{\varphi}_w$ in $n$ steps:
	\begin{itemize}
		\item At step 0, choose for each $\langle \alpha_i \rangle \varphi_i \in L(w)$ an event $e_i$ such 
		that $e_i \in \mathcal{I}(\alpha_i)$ and there exists a state $v_i$ with $w \relatedto{e_i} v_i$, and 
		$v_i, M \vDash \varphi_i$, and define $\displaystyle \mathcal{R}^0 = \bigcup_{e_i} \{w \relatedto{e_i} v_i \}$.
		\item At step $k+1$, let $v_1,...,v_m$ be the states $k$-reachable from $w$. For each of these states 
		proceed as was done for state $w$, and define a relation $R^k_{v_i}$, and then $\displaystyle \mathcal{R}^{k+1} = \bigcup_{i\leq m}R^{k}_{v_i}$ 
	\end{itemize}
	Finally, $R^{\varphi}_{w} = \bigcup_{k \leq n} R^k$.
   \item $\mathcal{P}^{\varphi}_{w} = \mathcal{P}$.
   \item $\mathcal{W}^{\varphi}_{w} = \{v \in \mathcal{W} \mid v \in \textsf{Dom}(\mathcal{R}^{\varphi}_w) \cup \textsf{Ran}(\mathcal{R}^{\varphi}_w)\}$.
   \item $I^{\varphi}_w(a_i) = \mathcal{I}(a_i)$, for every $e_i$. $\mathcal{I}^{\varphi}_{w}(p_i) = \mathcal{I}(p_i)$, for every $p_i \in \Phi_0$.
\end{itemize}
	Note that this model has an out-degree (the number of transitions coming out of any state) less than or equal to $\mathsf{D}_{\exists}(\varphi)$, since 
in each state of the new model, we only have one transition per existential subformula at the corresponding level.

	Suppose that $\textsf{d}(\varphi) = n$; we prove the result by induction. If $v$ is 
reachable in $n$ steps from $w$, then, by definition, $L(v)$ only contains propositional variables 
and deontic predicates, and therefore, by definition of $M^{\varphi}_w$, we have that 
$v, M^{\varphi}_{w} \vDash L(v)$.

	If $v$ is reachable in $k$ steps (with $k < n$) from $w$, then for each $\langle \alpha_i \rangle \varphi_i$ 
in $L(v)$ we have an $e_i \in \mathcal{I}^{\varphi}_w(\alpha_i)$ such that $v \relatedto{e_i} v'$; by induction 
we know that $v', M^{\varphi}_{w} \vDash \varphi_i$, and therefore $v, M^{\varphi}_{w} \vDash \langle \varphi_i \rangle \varphi_i$.

	Now, suppose that $\neg \langle \beta_j \rangle \psi_j \in L(v)$; we know that $\psi_j$ is a NF formula, and therefore 
$\psi_j = \psi'_1 \wedge ... \wedge \psi'_h$, and by definition of $\mathsf{SF}$, we have that $\neg \psi'_1,...,\neg \psi'_h \in \mathsf{SF}(\varphi, k+1)$.  
 Then, if for some $\psi'_l$ and state $v'$, we have $v', M \vDash \neg \psi'_{l}$ (where $v \relatedto{e_j} v'$ in $\mathcal{R}$ and $e_j \in \mathcal{I}(\beta_j)$), 
i.e., we have that $\neg \psi'_{l} \in L(v')$, then by induction we have $v', M^{\varphi}_{w} \vDash \neg \psi'_{l}$, which
implies that $v, M^{\varphi}_w \vDash \neg \langle \beta_j \rangle \psi_j$. This concludes the proof.
\end{theorem} 
	Note that $\bigwedge L_w(w) = \varphi$, and therefore we obtain the following corollary.
\begin{corollary} If $w, M \vDash \varphi$, then $w, M^{\varphi}_w \vDash \varphi$.
\end{corollary}
	Summarizing, given a model of a NF formula $\varphi$, we can build a new model with branching being at most $\mathsf{D}_{\exists}(\varphi)$. We are 
close to our original goal; using the model $M^{\varphi}_{w}$, we define another model over a restricted vocabulary that preserves property $\varphi$.
\begin{theorem}\label{theorem:reduced-model} Given a vocabulary $V = \langle \Delta_0, \Phi_0 \rangle$, a NF formula $\varphi$ such that $\Delta_0 > Pr(\varphi) + \mathsf{D}_{\exists}(\varphi)$, 
with $\mathsf{d}(\varphi)=n$ and a model $M$, if $w, M\vDash \varphi$, then there is a model $M^*$ over 
a vocabulary $V^*=\langle Pr(\varphi) \cup \{b_1,...,b_{\mathsf{D}_{\exists}(\varphi)} \}, \Phi_0 \rangle$ with  
the $b_i's$ being fresh action terms, such that $w, M^* \vDash \varphi$.\\
\textbf{Proof:}
First, given a model $M$ over 
a vocabulary $V=\langle \Delta_0, \Phi_0 \rangle$, we denote by $EQ(M)$ the set of equations true in $M$, and if we have a subset 
$S \subseteq \Delta_0$, we denote by $EQ^{S}(M)$ the set of equations built from primitive actions in $S$ which are true in $M$, i.e.,
\[
	EQ^S(M) = \{ \alpha =_{act} \beta \mid \alpha =_{act} \beta \in EQ(M) \wedge \alpha, \beta \in T_{BA}(S)\}
\]
where $T_{BA}(S)$ denotes the set of boolean terms built from variables in $S$.

	Suppose that $\mathsf{D}_{\exists}(\varphi)=c$. If $\# \Delta_0 > Pr(\varphi) + c$, then 
we define a model $M^{*} = \langle \mathcal{W}^*, \mathcal{R}^*, \mathcal{E}^*, \mathcal{I}^*, \mathcal{P}^* \rangle$ over 
the vocabulary $V^*=\langle \Delta^*_0 = Pr(\varphi) \cup \{b_1, \dots, b_c\}, \Phi_0 \rangle$, $b_1,...,b_c$ being fresh 
primitive actions. 
\begin{itemize}
	\item $\mathcal{E}^* = At(\Delta^*_0 / EQ^{Pr(\varphi)}(M))$.
	\item $\mathcal{W}^* = W^{\varphi}_{w}$.
	\item For each $v \in W^{\varphi}_w$ let $\{e^v_1,...,e^v_k\} \subseteq \mathcal{E}^{\varphi}_w$ be the set of events such that each 
		  $e^v_i$ satisfies either:	
		\begin{itemize}
			\item there exists a state $v_i$ and $v \relatedto{e_i} v_i \in \mathcal{R}^{\varphi}_w$, or
			\item there is a $\wper{\alpha_i} \in L(v)$ such that $e^{v}_{i} \in \mathcal{I}^{\varphi}_w(\alpha_i)$ and 
			$\mathcal{P}^{\varphi}_w(v,e^v_i)$, or
			\item there is a $\neg \per{\alpha_i} \in L(v)$ with  $e^{v}_{i} \in \mathcal{I}^{\varphi}_w(\alpha_i)$ and $(v, e^{v}_i) \notin \mathcal{P}^{\varphi}_w$
		\end{itemize}
		We know that $k \leq \mathsf{D}_{\exists}(\varphi)$, and then define for each such a $e^v_w$ a corresponding 
		event in $\mathcal{E}^*$ as follows:
		\[
			e^{v*}_i = (\bigsqcap_{a \in Pr(\varphi) \wedge e^v_i \in \mathcal{I}(a)} a) 
			\sqcap (\bigsqcap_{a' \in Pr(\varphi) \wedge e^v_i \notin \mathcal{I}(a')} \overline{a'})
			\sqcap (\bigsqcap_{b_j \in \Delta_0' \wedge b_j \neq b_i} \overline{b_j}) \sqcap b_i
		\]
		(where we use some enumeration of the fresh b's to determine each $b_i$); note that for these
		$e^{v*}_i$'s, we have: $e^v_i \in \mathcal{I}^{\varphi}_w(\alpha) \Leftrightarrow e^{v*}_i \in \mathcal{I}^{*}(\alpha)$, for 
		each $\alpha \in T_{BA}(Pr(\varphi))$. Now we use these $e^{v*}_i$'s to define:
		\begin{itemize}
			\item $\mathcal{R}^v = \{v \relatedto{e^{v*}_i} v_i \mid v \relatedto{e^{v}_i} v_i \in \mathcal{R}^{\varphi}_w\}$.
			\item $\mathcal{P}^v = \{(v, e^{v*}_i) \mid \mathcal{P}^{\varphi}_w(v, e^v_i) \}$.
		\end{itemize} 
		Using these sets defined for each state $v$, we define:
		$
			\mathcal{R}^* = \bigcup_{v \in \mathcal{W}^*} \mathcal{R}^v
		$
		and:
		\[
			\mathcal{P}^* = (\bigcup_{v \in \mathcal{W}^*} \mathcal{P}^v) 
							\cup \{ (v,e) \mid v \in \mathcal{W}^* \wedge \per{\alpha} \in L(v) \wedge e \in \mathcal{I}^*(\alpha) \}.
		\]
		\item Define $\mathcal{I}^*(a_i) = \{ [\gamma] \mid \vdash_{BA} \gamma \sqsubseteq a_i \}$, for every $a_i \in \Delta^*_0$.
		\item Define $\mathcal{I}^*(p_i) = \mathcal{I}^{\varphi}_w(p_i)$, for every atomic proposition $p_i$.
\end{itemize}
	Let us prove that this new model preserves properties of $L$.
	
	By the theorem above, we have that $w, M^{\varphi}_w \vDash \varphi$. As explained above, if 
we prove that $v, M^* \vDash L(v)$ for every $v$, we have that $w, M^* \vDash \varphi$. For the states reachable 
in $n$ steps, we have that $L(v)$ contains only propositional variables or deontic predicates; for the propositional 
predicates, the result is trivial. Now, suppose that $\per{\alpha} \in L(v)$, then $v, M^{\varphi}_{w} \vDash \per{\alpha}$, and 
then $v, M^* \vDash \per{\alpha}$, by the definition of $M^*$. If $\wper{\alpha} \in L(v)$, then $v, M^{\varphi}_w \vDash \wper{\alpha}$, 
and therefore there exists an $e_i \in \mathcal{I}^{\varphi}_w(\varphi)$ such that $\mathcal{P}^{\varphi}_w(v, e_i)$, but for this 
$e_i$ we have a corresponding $e^v_i$ such that $(v, e^v_i) \in \mathcal{P}^*$ and $e^v_i \in \mathcal{I}^*(\alpha)$, and therefore 
$v, M^* \vDash \wper{\alpha}$. If $\neg \per{\alpha} \in L(v)$, then we have an $e_i \in I^{\varphi}_w$ such that 
$\neg \mathcal{P}(v, e_i)$; for this $e_i$, we have an $e^v_i \in \mathcal{P}^*$ and by definition of $\mathcal{P}^*$ 
we have $\neg \mathcal{P}^*(v, e^v_i)$, since $\neg \mathcal{P}^{\varphi}_w(v, e_i)$ and 
$\per{\alpha} \notin L(v)$, otherwise $L(v)$ is inconsistent. Therefore, $v, M^* \vDash \neg \per{\alpha}$. 
	If $\neg \wper{\alpha} \in L(v)$, then we have have $\neg \mathcal{P}^{\varphi}_w(e_i, v)$ for every $e_i \in \mathcal{I}^*(\alpha)$;  
if we have $\per{\alpha} \in L(v)$, then $\alpha =_{act} \emptyset$, an equation which is also true in $M^*$ and therefore 
$v, M^* \vDash \neg \wper{\alpha}$. If $\alpha \neq_{act} \emptyset$, then $\per{\alpha} \notin L(v)$, and there is 
no way to introduce a tuple $(v, e^v_i)$ in $\mathcal{P}^*$, so $v, M^* \vDash \neg \wper{\alpha}$.

	Now, suppose that $v$ is reachable in $k < n$ steps from $w$; for the deontic predicates and propositional variables 
the proof proceeds as before. If $\langle \alpha_i \rangle \varphi_i \in L(v)$,   then $v, M^{\varphi}_w \vDash \langle \alpha_i \rangle \varphi_i$, and  
so there is an $e_i \in \mathcal{I}^{\varphi}_w(\alpha)$ such that $v \relatedto{e_i} v_i$ and $v_i, M^{\varphi}_w \vDash \varphi_i$. Using 
induction we get $v_i, M^* \vDash \varphi_i$, and we have, by definition of $M^*$, that $v \relatedto{e^v_i} v_i$; this implies that
$v, M^* \vDash \langle \alpha_i \rangle \varphi_i$. If $\neg \langle \beta_i \rangle \psi_i \in L(v)$, then $v, M^{\varphi}_w \vDash \psi_i$, 
which means that for all $e_i$ such that $e_i \in \mathcal{I}^{\varphi}_w(\beta_i)$ and $v \relatedto{e_i} v_i$, we have 
$v_i, M^{\varphi}_w \vDash \neg \psi$. Since $\psi_i$ is a NF formula, it is a conjunction of formulae, i.e., 
$\psi_i = \psi^1_i \wedge \dots \wedge \psi^m_i$, and, for some of these $\psi_i$'s, we have 
$v, M^{\varphi}_w \vDash \neg \psi^j_i$, and by definition of $L$ we have $\neg \psi^j_i \in L(v')$ and therefore, by induction, 
$v', M^* \vDash \neg \psi^j_i$, which implies $w, M^* \vDash \neg \langle \beta_i \rangle \psi_i$. The theorem follows.
\end{theorem}
	Thus, $\textsf{D}_{\exists}(\varphi)$ gives us a bound for the number of new primitive symbols that we need 
to verify a given formula. From this theorem we get the following corollary:
\begin{corollary}\label{theorem:bound} For any DNF formula $\varphi$ with $\mathsf{D}_{\exists}(\varphi) = n$, if we have a vocabulary 
$V = \langle \Delta_0, \Phi_0 \rangle$ and a model $M$ of $V$ such that $w, M \vDash \varphi$, then there exists 
a model $M'$ of a vocabulary $V' = \langle Pr(\varphi) \cup \{b_1,...,b_k\}, \Phi_0 \rangle$ such that 
$w', M' \vDash \varphi$ and $k \leq n$.\\
\textbf{Proof.}
Suppose that $w,M \vDash \varphi$ for some $M$ over a vocabulary $V$. Since $\varphi$ is 
in $DNF$, we know that $\varphi = \varphi_1 \vee ... \vee \varphi_m$ (each $\varphi_i$ being a NF formulae), and 
therefore $w, M \vDash \varphi_i$ for some $i$. By theorem \ref{theorem:reduced-model} we know that there exists a model $M'$ of a vocabulary 
$V' = \langle Pr(\varphi) \cup \{b_1,...,b_k \}, \Phi_0 \rangle$ with $k = \mathsf{D}_{\exists}(\varphi_i) \leq \mathsf{D}_{\exists}(\varphi)$ 
such that $w', M' \vDash \varphi_i$ and then $w', M' \vDash \varphi$. (Note that we can ensure that each $Pr(\varphi) = Pr(\varphi_i)$, by adding 
the formulae $[a_1 \sqcup ... \sqcup a_t] \top$ to each $\varphi_i$ with $Pr(\varphi) = \{a_1,...,a_t\}$, these formulae do not modify 
the truth value of the former.)
\end{corollary}
        Roughly speaking, this theorem says that, if we cannot get a model with $n$  (with $\mathsf{D}_{\exists}(\varphi) = n$) new primitive actions, we will not get a model by adding 
further primitive actions to the language.       
        Because each formula is equivalent to a DNF formula, the above result gives us a 
bound for checking every formula (where the bound depends on the formula under consideration). 

	The method is as follows: given a formula $\varphi$, take its negation and
then develop a tableau taking into account at most $\mathsf{D}_{\exists}(\neg \varphi)=n$  (which can be calculated for any formula, as explained above)
primitive actions; if the tableau is closed, then the formula $\varphi$ is valid for any extension
of its vocabulary. 
	It is worth noting that we have two kinds of validities: we have formulae which are valid with respect to 
one vocabulary (i.e., these formulae are true with respect to all the models of this vocabulary). We can call this notion of validity  
\textit{local validity}. And we have formulae which are valid with respect to every vocabulary, i.e., a \textit{global validity}. 
	For example the formula $[a \sqcup b] \varphi \leftrightarrow [\univ] \varphi$ is valid in the vocabulary 
$\langle \{a, b\}, \{p, q, s, \dots \} \rangle$ but it is not valid in the vocabulary $\langle \{a, b, c\}, \{p, q, s, \dots \} \rangle$.     

\section{Time Complexity} \label{section:complexity}
	We present some results about the time complexity  of the method presented above. Our first theorem says that checking local satisfiability is 
polynomial w.r.t. space.
\begin{theorem}\label{theorem:local-complexity}
	Checking local satisfiability with the tableaux for DPL is PSPACE.\\
\textbf{Proof.} Let us note that, given a vocabulary, the 
number of its actions is fixed, and then, the number of atomic action terms is constant with respect to formula length. 
Since the branching is bounded by the number of atoms, it is bounded by a constant, and the length of any branch 
is linearly bounded by the length of the formula.  That is, it is straightforward, given a formula $\varphi$, to write a procedure that 
inspects each branch in time $O(a^{|\varphi|})$, where $|\varphi|$ is the length of $\varphi$ and $a$ is the number of atomic action terms of the given 
vocabulary. Furthermore, since we only need to inspect a branch at a time, we can develop an algorithm that behaves in a backtracking 
way; it only needs to keep a record of the current branch, each atom can be described by a binary number of constant length w.r.t. the 
formula length, and (in each branch) we only use an atom for each existential subformula; i.e.,  
since this branch is linearly bounded by the length of the formula (the number of subformulae of a formula is linear w.r.t. the length of the formula), 
we only need polynomial space to develop the algorithm.
\end{theorem} 
	However, when we want to check if a formula is valid in any possible extension of the vocabulary, we need to 
use the the results shown in section \ref{section:language-extension}. Interestingly, in this case we also obtain that the method is polynomial w.r.t space.
\begin{theorem}\label{theorem:global-complexity}
	Checking global satisfiability with the tableaux for DPL is PSPACE.\\
\textbf{Proof.}
Let us analyze the complexity of the global satisfiability, i.e., when we want to check the satisfiability of a formula in 
any vocabulary extension. In this case, we add a number of actions to a vocabulary, this number 
is bounded by the length of the formula. Note that this does not affect the length of  branches since we still use one 
action atom for each existential subformula appearing in the original formula. Moreover, each atom appearing in the branch can 
be represented using a binary number of a length linearly bounded by the length of the formula (we have a part of the binary number which 
is constant since it represents the occurrence of the original action, plus a binary number that represents the new actions which 
are bounded by the length of the formula). And therefore using the same procedure as before, but taking into account the new actions we obtain a PSPACE 
algorithm.
\end{theorem} 
	Let us compare the complexity of the tableaux method developed here with the complexity of related logics. 
	Boolean modal logic is NExpTime-Complete \cite{Lutz2000} since the global modality allows us to encode 
complex problems. Restricting BML to signatures with only a finite number of relation symbols is 
ExpTime-Complete \cite{Lutz2000}. That is, the complexity of our tableaux is 
therefore relatively acceptable w.r.t. the complexity of well-known logics. 
	Taking into account that the modal logic \textbf{K} 
is PSPACE-Complete, we can expect that the complexity of the tableaux is aligned to the complexity of deciding validity in 
DPL. 

\begin{figure}
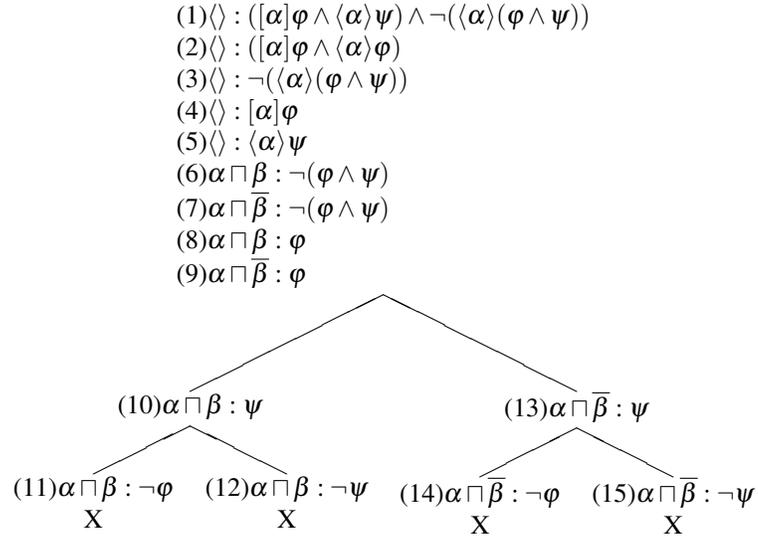

\centering
\small{
\Tree[.{\begin{tabular}{l}(1)$\langle \rangle: ([\alpha]\varphi \wedge \langle \alpha \rangle \psi) \wedge \neg (\langle \alpha \rangle (\varphi \wedge \psi))$\\
		 (2)$\langle \rangle : ([\alpha]\varphi \wedge \langle \alpha \rangle \varphi)$\\ 
         (3)$\langle \rangle : \neg (\langle \alpha \rangle (\varphi \wedge \psi))$\\
         (4)$\langle \rangle : [\alpha]\varphi$\\
         (5)$\langle \rangle : \langle \alpha \rangle \psi$\\
         (6)$\alpha \sqcap \beta : \neg (\varphi \wedge \psi)$\\
         (7)$\alpha \sqcap \overline{\beta} : \neg (\varphi \wedge \psi)$\\
         (8)$\alpha \sqcap \beta : \varphi $\\
         (9)$\alpha \sqcap \overline{\beta} : \varphi$
		\end{tabular}}        
        [.{(10)$\alpha \sqcap \beta: \psi$} [.{(11)$\alpha \sqcap \beta : \neg \varphi$ \\ X} ] [.{(12)$\alpha \sqcap \beta:\neg \psi$ \\ X} ] ]
        [.{(13)$\alpha \sqcap \overline{\beta}:\psi$} [.{(14)$\alpha \sqcap \overline{\beta}:\neg \varphi$ \\ X } ] [.{(15)$\alpha \sqcap \overline{\beta}:\neg \psi$ \\ X} ] ] ]
}
\caption{Tableau for $([\alpha]\varphi \wedge \langle \alpha \rangle \psi) \rightarrow \langle \alpha \rangle (\varphi \wedge \psi)$}\label{fig:example1}
\end{figure}

	Now we give some examples. In figure \ref{fig:example1} we build the tableau for the formula: 
$([\alpha]\varphi \wedge \langle \alpha \rangle \psi) \rightarrow \langle \alpha \rangle (\varphi \wedge \psi)$.
This formula is one of the axioms given for dynamic logic in \cite{DynamicLogic}. The crosses at the end of each branch
mean that those branches are closed. Note that here we are using a new action symbol.

	 We have added numbers in the formulae to better explain the example. Formula (1) is the negation of the formula to be proven. 
Formulae (2) and (3) are obtained by applying the rule $A$ to the negation of the implication, 
formulae (4) and (5) are obtained from formula (2) using the $A$ rule. Formulae (6) and (7) follow from formula (3) by application of the 
$N$ rule. In a similar way, we obtained formulae (8) and (9) from formula 4. After that,
we have branching using the $P$ rule. Finally, we apply the $B$ rule to formulae (6) and (7) and we obtain the leaves 
closing the tableau.

\begin{figure}
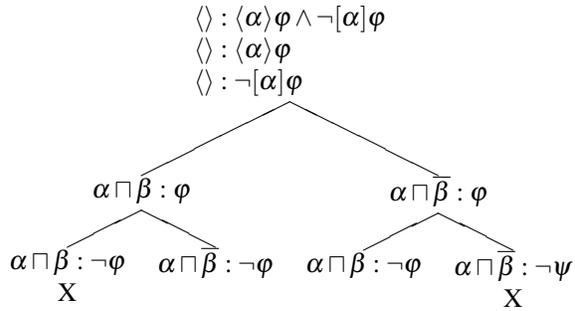

\centering
\small{
\Tree [.{\begin{tabular}{l}$\langle \rangle: \langle \alpha \rangle \varphi \wedge \neg [\alpha]\varphi$\\
		 $\langle \rangle : \langle \alpha \rangle \varphi$\\ 
         $\langle \rangle : \neg [\alpha] \varphi$\end{tabular}}         
        [.{$\alpha \sqcap \beta: \varphi$} [.{$\alpha \sqcap \beta : \neg \varphi$ \\ X} ] [.{$\alpha \sqcap \overline{\beta}:\neg \varphi$} ] ]
        [.{$\alpha \sqcap \overline{\beta}:\varphi$} [.{$\alpha \sqcap \beta:\neg \varphi$} ] [.{$\alpha \sqcap \overline{\beta}:\neg \psi$ \\ X} ] ] ]
}
\caption{Tableau for $\langle \alpha \rangle \varphi \rightarrow [ \alpha ] \varphi$}\label{fig:example2}
\end{figure}

\begin{figure}
\centering
\small{
\[
\xymatrix{                  & & \bullet w' \vDash \varphi \\
          w \bullet \ar[urr]^{\alpha \sqcap \beta} \ar[drr]^{\alpha \sqcap \overline{\beta}} & &  \\
                            & & \bullet w'' \vDash \neg \varphi\\
}
\]
}
\caption{Counterexample for $\langle \alpha \rangle \varphi \rightarrow [ \alpha ] \varphi$}\label{fig:counterexample1}
\end{figure}
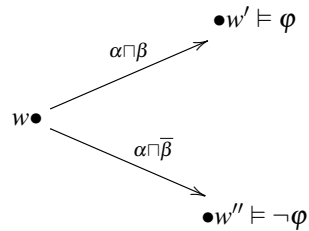

 Now, consider the following formula (which is not valid): $\langle \alpha \rangle \varphi \rightarrow [ \alpha ] \varphi$.  
The tableau for it is shown in figure \ref{fig:example2}.
Note that, in this case, we use a new action symbol (following corollary \ref{theorem:bound}). First, we reduce the implication. After that we use the rule $P$ on the second formula, and then we use rule $P$
again in the third formula. We can observe that this tableau has some open branches and using them we can build a ``counterexample'' (shown 
in figure \ref{fig:counterexample1}).
        Note that we can use the labels in the formulae to put the labels on the transitions, indicating in this
way which actions were executed and which were not.

\section{Further Remarks}
	In this paper we have investigated further the properties of the tableaux system presented in \cite{DEON08}. One of 
the main features of the system is that it uses the underlying algebra of actions to produce tableaux, enabling it to manage 
successfully the intersection and complement operators on actions. Moreover, the algebra of actions allows 
us to extend the propositional tableaux system to manage temporal predicates. 

	A relevant point demonstrated in the paper is that, though we have a finite number of actions, it is possible to  prove properties 
which are valid in any extension of the actual vocabulary, which seems to be very useful in practice to verify computing components which 
could be part of bigger systems. This kind of completeness with respect to language extension is also preserved for the temporal 
version of the logic. In addition, we have investigated the time complexity of the tableaux system and we prove that it is in PSPACE 
in comparison with related logics, which are in EXPTIME.

	This deontic logic was presented in \cite{JAL2009}, where several of its properties, and some examples of 
application, were shown. In those papers we proposed this logic to specify and verify properties related to
fault-tolerance. It seems very useful to apply the tableaux system described here to such examples. We leave this as further work.


\bibliographystyle{eptcs}
\bibliography{../biblio/biblio}

\begin{thebibliography}{10}
\providecommand{\bibitemdeclare}[2]{}
\providecommand{\surnamestart}{}
\providecommand{\surnameend}{}
\providecommand{\urlprefix}{Available at }
\providecommand{\url}[1]{\texttt{#1}}
\providecommand{\href}[2]{\texttt{#2}}
\providecommand{\urlalt}[2]{\href{#1}{#2}}
\providecommand{\doi}[1]{doi:\urlalt{http://dx.doi.org/#1}{#1}}
\providecommand{\bibinfo}[2]{#2}

\bibitemdeclare{book}{Blackburn}
\bibitem{Blackburn}
\bibinfo{author}{P.~\surnamestart Blackburn\surnameend}, \bibinfo{author}{M.de
  \surnamestart Rijke\surnameend} \& \bibinfo{author}{Y.de \surnamestart
  Venema\surnameend} (\bibinfo{year}{2001}): \emph{\bibinfo{title}{Modal
  Logic}}.
\newblock \bibinfo{publisher}{Cambridge Tracts in Theoretical Computer Science
  53}.

\bibitemdeclare{phdthesis}{BroersenThesis}
\bibitem{BroersenThesis}
\bibinfo{author}{J.~\surnamestart Broersen\surnameend} (\bibinfo{year}{2003}):
  \emph{\bibinfo{title}{Modal Action Logics for Reasoning about Reactive
  Systems}}.
\newblock Ph.D. thesis, \bibinfo{school}{Vrije University}.

\bibitemdeclare{phdthesis}{PhDThesis}
\bibitem{PhDThesis}
\bibinfo{author}{Pablo~F. \surnamestart Castro\surnameend}
  (\bibinfo{year}{2009}): \emph{\bibinfo{title}{Deontic Action Logics for the
  Specification and Analysis of Fault-Tolerance}}.
\newblock Ph.D. thesis, \bibinfo{school}{McMaster University, Department of
  Computing and Software}.

\bibitemdeclare{conference}{DEON08}
\bibitem{DEON08}
\bibinfo{author}{Pablo~F. \surnamestart Castro\surnameend} \&
  \bibinfo{author}{T.S.E. \surnamestart Maibaum\surnameend}
  (\bibinfo{year}{2008}): \emph{\bibinfo{title}{A Tableaux System for Deontic
  Action Logic}}.
\newblock In: {\sl \bibinfo{booktitle}{Proceedings of 9th International
  Conference on Deontic Logic in Computer Science , Luxembourg.}},
  \bibinfo{publisher}{Springer-Verlag}, \doi{10.1007/978-3-540-70525-3\_4}.

\bibitemdeclare{article}{JAL2009}
\bibitem{JAL2009}
\bibinfo{author}{Pablo~F. \surnamestart Castro\surnameend} \&
  \bibinfo{author}{T.S.E. \surnamestart Maibaum\surnameend}
  (\bibinfo{year}{2009}): \emph{\bibinfo{title}{Deontic Action Logic, Atomic
  Boolean Algebra and Fault-Tolerance}}.
\newblock {\sl \bibinfo{journal}{Journal of Applied Logic}}
  \bibinfo{volume}{7}(\bibinfo{number}{4}), pp. \bibinfo{pages}{441--466},
  \doi{10.1016/j.jal.2009.02.001}.

\bibitemdeclare{book}{Chellas}
\bibitem{Chellas}
\bibinfo{author}{Brian~F. \surnamestart Chellas\surnameend}
  (\bibinfo{year}{1999}): \emph{\bibinfo{title}{Modal Logic: An Introduction}}.
\newblock \bibinfo{publisher}{Cambridge University Press}.

\bibitemdeclare{conference}{EmersonSurvey}
\bibitem{EmersonSurvey}
\bibinfo{author}{E.A. \surnamestart Emerson\surnameend} (\bibinfo{year}{1995}):
  \emph{\bibinfo{title}{Temporal and Modal Logic}}.
\newblock In: {\sl \bibinfo{booktitle}{Handbook of Theoretical Computer
  Science}}, \bibinfo{publisher}{Elsevier}, pp. \bibinfo{pages}{995--1072}.

\bibitemdeclare{conference}{Fine75}
\bibitem{Fine75}
\bibinfo{author}{Kit \surnamestart Fine\surnameend} (\bibinfo{year}{1975}):
  \emph{\bibinfo{title}{Normal Forms in Modal Logics}}.
\newblock In: {\sl \bibinfo{booktitle}{Notre Dame Journal of Formal Logic}},
  \bibinfo{volume}{XVI}, pp. \bibinfo{pages}{229--237},
  \doi{10.1305/ndjfl/1093891703}.

\bibitemdeclare{book}{Fitting}
\bibitem{Fitting}
\bibinfo{author}{M.~\surnamestart Fitting\surnameend} (\bibinfo{year}{1990}):
  \emph{\bibinfo{title}{First-Order Logic and Automated Theorem Proving}}.
\newblock \bibinfo{publisher}{Springer-Verlag},
  \doi{10.1007/978-1-4684-0357-2}.

\bibitemdeclare{conference}{Fitting72}
\bibitem{Fitting72}
\bibinfo{author}{M.~\surnamestart Fitting\surnameend} (\bibinfo{year}{April
  1972}): \emph{\bibinfo{title}{Tableau Methods of Proof for Modal Logics}}.
\newblock In: {\sl \bibinfo{booktitle}{Notre Dame Journal of Formal Logic}},
  \bibinfo{volume}{XIII}, \doi{10.1305/ndjfl/1093894722}.

\bibitemdeclare{inbook}{Gabbay96}
\bibitem{Gabbay96}
\bibinfo{author}{Dov~M. \surnamestart Gabbay\surnameend}
  (\bibinfo{year}{1996}): \emph{\bibinfo{title}{Labelled Deductive Systems,
  Volume 1}}.
\newblock \bibinfo{publisher}{Oxford University Press}.

\bibitemdeclare{inproceedings}{BooleanModalLogic}
\bibitem{BooleanModalLogic}
\bibinfo{author}{G.~\surnamestart Gargov\surnameend} \&
  \bibinfo{author}{S.~\surnamestart Passy\surnameend} (\bibinfo{year}{1990}):
  \emph{\bibinfo{title}{A Note on Boolean Logic}}.
\newblock In \bibinfo{editor}{\surnamestart P.P.Petkov\surnameend}, editor:
  {\sl \bibinfo{booktitle}{Proceedings of the Heyting Summerschool}},
  \bibinfo{publisher}{Plenum Press}, \doi{10.1007/978-1-4613-0609-2\_21}.

\bibitemdeclare{conference}{Giacomo-Massacci}
\bibitem{Giacomo-Massacci}
\bibinfo{author}{G.~\surnamestart Giacomo\surnameend} \&
  \bibinfo{author}{F.~\surnamestart Massacci\surnameend}
  (\bibinfo{year}{1996}): \emph{\bibinfo{title}{Tableaux and Algorithms for
  Propositional Dynamic Logic with Converse}}.
\newblock In: {\sl \bibinfo{booktitle}{Conference on Automated Deduction}},
  \doi{10.1006/inco.1999.2852}.

\bibitemdeclare{techreport}{Gore95}
\bibitem{Gore95}
\bibinfo{author}{Rajeev \surnamestart Gor\'e\surnameend}
  (\bibinfo{year}{1995}): \emph{\bibinfo{title}{Tableau Methods for Modal and
  Temporal Logics}}.
\newblock \bibinfo{type}{Technical Report} \bibinfo{number}{TR-ARP-15-95},
  \bibinfo{institution}{Australian National University}.

\bibitemdeclare{inbook}{DynamicLogic}
\bibitem{DynamicLogic}
\bibinfo{author}{D.~\surnamestart Harel\surnameend},
  \bibinfo{author}{D.~\surnamestart Kozen\surnameend} \&
  \bibinfo{author}{J.~\surnamestart Tiuryn\surnameend} (\bibinfo{year}{2000}):
  \emph{\bibinfo{title}{Dynamic Logic}}.
\newblock \bibinfo{publisher}{MIT Press}.

\bibitemdeclare{inproceedings}{MaibaumKhosla}
\bibitem{MaibaumKhosla}
\bibinfo{author}{S.~\surnamestart Khosla\surnameend} \& \bibinfo{author}{T.S.E.
  \surnamestart Maibaum\surnameend} (\bibinfo{year}{1985}):
  \emph{\bibinfo{title}{The Prescription and Description of State-Based
  Systems.}}
\newblock In \bibinfo{editor}{H.Barringer \surnamestart B.Banieqnal\surnameend}
  \& \bibinfo{editor}{\surnamestart A.Pnueli\surnameend}, editors: {\sl
  \bibinfo{booktitle}{Temporal Logic in Computation}},
  \bibinfo{publisher}{Springer-Verlag}, \doi{10.1007/3-540-51803-7\_30}.

\bibitemdeclare{inproceedings}{Lutz2000}
\bibitem{Lutz2000}
\bibinfo{author}{Carsten \surnamestart Lutz\surnameend} \&
  \bibinfo{author}{Ulrike \surnamestart Sattler\surnameend}
  (\bibinfo{year}{2000}): \emph{\bibinfo{title}{The Complexity of Reasoning
  with Boolean Modal Logics}}.
\newblock In: {\sl \bibinfo{booktitle}{Advances in Modal Logic 3}},
  \bibinfo{publisher}{World Scientific}, pp. \bibinfo{pages}{329--348},
  \doi{10.1142/9789812776471\_0018}

\bibitemdeclare{article}{Massacci2000}
\bibitem{Massacci2000}
\bibinfo{author}{Fabio \surnamestart Massacci\surnameend}
  (\bibinfo{year}{2000}): \emph{\bibinfo{title}{Single Step Tableaux for Modal
  Logics}}.
\newblock {\sl \bibinfo{journal}{J. Autom. Reasoning}}
  \bibinfo{volume}{24}(\bibinfo{number}{3}), pp. \bibinfo{pages}{319--364},
  \doi{10.1023/A:1006155811656}.

\bibitemdeclare{inproceedings}{J.J.Meyer}
\bibitem{J.J.Meyer}
\bibinfo{author}{J.J. \surnamestart Meyer\surnameend} (\bibinfo{year}{1988}):
  \emph{\bibinfo{title}{A Different Approach to Deontic Logic: Deontic Logic
  Viewed as Variant of Dynamic Logic}}.
\newblock In: {\sl \bibinfo{booktitle}{Notre Dame Journal of Formal Logic}},
  \bibinfo{volume}{29}, \doi{10.1305/ndjfl/1093637776}.

\bibitemdeclare{book}{Monk}
\bibitem{Monk}
\bibinfo{author}{J.D. \surnamestart Monk\surnameend} (\bibinfo{year}{1976}):
  \emph{\bibinfo{title}{Mathematical Logic}}.
\newblock \bibinfo{series}{Graduate Texts in Mathematics},
  \bibinfo{publisher}{Springer-Verlag}.

\bibitemdeclare{book}{Pratt78}
\bibitem{Pratt78}
\bibinfo{author}{V.R. \surnamestart Pratt\surnameend} (\bibinfo{year}{1978}):
  \emph{\bibinfo{title}{A Practical Decision Method for Propositional Dynamic
  Logic}}.
\newblock \bibinfo{publisher}{ACM Symposium on Theory of Computing},
  \doi{10.1145/800133.804362}.

\bibitemdeclare{article}{Segerberg82}
\bibitem{Segerberg82}
\bibinfo{author}{Krister \surnamestart Segerberg\surnameend}
  (\bibinfo{year}{1982}): \emph{\bibinfo{title}{A Deontic Logic of Action}}.
\newblock {\sl \bibinfo{journal}{Studia Logica}} \bibinfo{volume}{41}, pp.
  \bibinfo{pages}{269--282}, \doi{10.1007/BF00370348}.

\bibitemdeclare{book}{Smullyan68}
\bibitem{Smullyan68}
\bibinfo{author}{R.M. \surnamestart Smullyan\surnameend}
  (\bibinfo{year}{1968}): \emph{\bibinfo{title}{First-Order Logic}}.
\newblock \bibinfo{publisher}{Springer-Verlag New York},
  \doi{10.1007/978-3-642-86718-7}.

\bibitemdeclare{conference}{TrypuzKulicki2010}
\bibitem{TrypuzKulicki2010}
\bibinfo{author}{Robert \surnamestart Trypuz\surnameend} \&
  \bibinfo{author}{Piotr \surnamestart Kulicki\surnameend}
  (\bibinfo{year}{2010}): \emph{\bibinfo{title}{Towards Metalogical
  Systematisation of Deontic Action Logics Based on Boolean Algebra}}.
\newblock In: {\sl \bibinfo{booktitle}{Deontic Logic in Computer Science, 10th
  International Conference}}, \bibinfo{publisher}{Lecture Notes in Computer
  Science 6181 Springer}, \doi{10.1007/978-3-642-14183-6\_11}.

\end{thebibliography}
%
\end{document}